\documentstyle[12pt]{article}

\begin{document}

\author{C. Bizdadea\thanks{%
e-mail addresses: bizdadea@central.ucv.ro and bizdadea@hotmail.com}, L.
Saliu, S. O. Saliu\thanks{%
e-mail addresses: osaliu@central.ucv.ro and odile\_saliu@hotmail.com} \\
Department of Physics, University of Craiova\\
13 A. I. Cuza Str., Craiova RO-1100, Romania}
\title{On Chapline-Manton couplings: a cohomological approach }
\maketitle

\begin{abstract}
Chern-Simons couplings between Yang-Mills gauge fields and an abelian 2-form
are derived by means of cohomological arguments.

PACS number: 11.10.Ef
\end{abstract}

\section{Introduction}

The problem of constructing consistent interactions among fields with gauge
freedom \cite{1}--\cite{4} has been reformulated \cite{5} in the framework
of the antifield BRST formalism \cite{6}--\cite{10} as a problem of
consistent deformation of the master equation in the sense of deformation
theory \cite{11}. This technique has been applied to Chern-Simons models 
\cite{5}, Yang-Mills theories \cite{12}, and two-form gauge fields \cite{13}.

In this paper we investigate another interesting interaction, namely, the
consistent interaction between the Yang-Mills vector potential and an
abelian 2-form. As it will be seen, the procedure to be developed leads to
combined Yang-Mills-2-form system coupled through a Yang-Mills Chern-Simons
term (the Chapline-Manton model) \cite{14}--\cite{17}. Chern-Simons
couplings of a two-form to Yang-Mills gauge fields play a crucial role in
the Green-Schwarz anomaly cancellation mechanism \cite{18} and, therefore,
are important in string theory \cite{19}.

The strategy employed in this paper is the following. We start with an
action describing pure Yang-Mills theory and an abelian 2-form (the ``free''
theory), and determine the BRST differential $s$ of this uncoupled model,
which can be written as the sum between the Koszul-Tate differential and the
exterior longitudinal derivative along the gauge orbits, $s=\delta +\gamma $%
. Next, we deform the solution to the master equation of the uncoupled
model. The first-order deformation belongs to $H^0\left( s|d\right) $, where 
$d$ is the exterior space-time derivative. As $s=\delta +\gamma $, the
computation of the cohomology $H^0\left( s|d\right) $ proceeds by expanding
the co-cycles according to the antighost number. Subsequently, we derive the
second-order deformation, and thus, the solution to the master equation of
the interacting theory (the higher-order deformation terms vanish). The
antighost number zero piece in the deformed solution is nothing but the
action of the Chapline-Manton model, while the terms linear in the
antifields of the original fields and ghosts indicate that the added
interactions deform the gauge transformations, as well as the gauge algebra.
In this way, our approach emphasizes a complex consistent interaction that
deforms both the gauge transformations and their algebra.

\section{BRST symmetry for the ``free'' theory}

In this section we derive the BRST symmetry for the ``free'' theory. In this
respect, we begin with a Lagrangian action equal with the sum between the
actions of Yang-Mills theory and a free 2-form 
\begin{equation}
\label{1}S_0^L\left[ A_\mu ^a,B_{\mu \nu }\right] =\int d^Dx\left( -\frac
14F_{\mu \nu }^aF_a^{\mu \nu }-\frac 1{12}F_{\mu \nu \rho }F^{\mu \nu \rho
}\right) , 
\end{equation}
where 
\begin{equation}
\label{2}F_{\mu \nu }^a=\partial _\mu A_\nu ^a-\partial _\nu A_\mu
^a-f_{\;\;bc}^aA_\mu ^bA_\nu ^c, 
\end{equation}
\begin{equation}
\label{3}F_{\mu \nu \rho }=\partial _\mu B_{\nu \rho }+\partial _\rho B_{\mu
\nu }+\partial _\nu B_{\rho \mu }\equiv \partial _{\left[ \mu \right.
}B_{\left. \nu \rho \right] }. 
\end{equation}
Action (\ref{1}) is invariant under the gauge transformations 
\begin{equation}
\label{4}\delta _\epsilon A_\mu ^a=\left( D_\mu \right) _{\;\;b}^a\epsilon
^b,\;\;\delta _\epsilon B_{\mu \nu }=\partial _\mu \epsilon _\nu -\partial
_\nu \epsilon _\mu \equiv \partial _{\left[ \mu \right. }\epsilon _{\left.
\nu \right] }, 
\end{equation}
with $\left( D_\mu \right) _{\;\;b}^a=\delta _{\;\;b}^a\partial _\mu
+f_{\;\;bc}^aA_\mu ^c$ the covariant derivative. The gauge transformations
of the 2-form are first-stage reducible as they vanish for $\epsilon _\mu
=\partial _\mu \epsilon $. Consequently, the solution to the master equation
of the ``free'' theory reads as%
\begin{eqnarray}\label{5}
& &S=S_0^L\left[ A_\mu ^a,B_{\mu \nu }\right] +\int d^Dx\left( A_a^{*\mu
}\left( D_\mu \right) _{\;\;b}^a\eta ^b-\frac 12f_{\;\;bc}^a\eta _a^{*}\eta
^b\eta ^c+\right. \nonumber \\
& &\left. B^{*\mu \nu }\partial _{\left[ \mu \right. }\eta _{\left.
\nu \right] }+\eta ^{*\mu }\partial _\mu \eta \right) . 
\end{eqnarray}
In (\ref{5}), $\eta ^a$ and $\eta _\mu $ denote the fermionic ghosts with
ghost number one, while $\eta $ represents the bosonic ghost for ghost of
ghost number two that appears due to the reducibility. The variables $%
A_a^{*\mu }$, $B^{*\mu \nu }$, $\eta _a^{*}$, $\eta ^{*\mu }$ and $\eta ^{*}$
stand for the antifields. The first two sets of antifields have ghost number
minus one, the next two sets display ghost number minus two, and the last
possesses ghost number minus three. The ghost number is defined like the
difference between pure ghost number ($pgh$) and antighost number ($antigh$%
), where 
\begin{equation}
\label{6}pgh\left( A_\mu ^a\right) =0,\;pgh\left( B_{\mu \nu }\right)
=0,\;pgh\left( A_a^{*\mu }\right) =0, 
\end{equation}
\begin{equation}
\label{7}pgh\left( B^{*\mu \nu }\right) =0,\;pgh\left( \eta _a^{*}\right)
=0,\;pgh\left( \eta ^{*\mu }\right) =0,\;pgh\left( \eta ^{*}\right) =0, 
\end{equation}
\begin{equation}
\label{8}pgh\left( \eta ^a\right) =1,\;pgh\left( \eta _\mu \right)
=1,\;pgh\left( \eta \right) =2, 
\end{equation}
\begin{equation}
\label{9}antigh\left( A_\mu ^a\right) =0,\;antigh\left( B_{\mu \nu }\right)
=0,\;antigh\left( A_a^{*\mu }\right) =1, 
\end{equation}
\begin{equation}
\label{10}antigh\left( B^{*\mu \nu }\right) =1,\;antigh\left( \eta
_a^{*}\right) =2,\;antigh\left( \eta ^{*\mu }\right) =2, 
\end{equation}
\begin{equation}
\label{11}antigh\left( \eta ^{*}\right) =3,\;antigh\left( \eta ^a\right)
=0,\;antigh\left( \eta _\mu \right) =0,\;antigh\left( \eta \right) =0. 
\end{equation}
The BRST differential $s\bullet =\left( \bullet ,S\right) $ of the ``free''
theory splits as 
\begin{equation}
\label{12}s=\delta +\gamma , 
\end{equation}
where $\delta $ is the Koszul-Tate differential and $\gamma $ represents the
longitudinal exterior derivative along the gauge orbits. The symbol $\left(
,\right) $ is the usual notation for antibracket in the antifield formalism.
Thus, we have 
\begin{equation}
\label{13}\delta A_\mu ^a=0,\;\gamma A_\mu ^a=\left( D_\mu \right)
_{\;\;b}^a\eta ^b, 
\end{equation}
\begin{equation}
\label{14}\delta B_{\mu \nu }=0,\;\gamma B_{\mu \nu }=\partial _{\left[ \mu
\right. }\eta _{\left. \nu \right] }, 
\end{equation}
\begin{equation}
\label{15}\delta \eta ^a=0,\;\gamma \eta ^a=-\frac 12f_{\;\;bc}^a\eta ^b\eta
^c, 
\end{equation}
\begin{equation}
\label{16}\delta \eta _\mu =0,\;\gamma \eta _\mu =\partial _\mu \eta , 
\end{equation}
\begin{equation}
\label{17}\delta \eta =0,\;\gamma \eta =0, 
\end{equation}
\begin{equation}
\label{18}\delta A_a^{*\mu }=-\left( D_\nu \right) _a^{\;\;b}F_b^{\nu \mu
},\;\gamma A_a^{*\mu }=f_{\;\;ac}^bA_b^{*\mu }\eta ^c, 
\end{equation}
\begin{equation}
\label{19}\delta B^{*\mu \nu }=-\frac 12\partial _\rho F^{\rho \mu \nu
},\;\gamma B^{*\mu \nu }=0, 
\end{equation}
\begin{equation}
\label{20}\delta \eta _a^{*}=\left( D_\mu \right) _a^{\;\;b}A_b^{*\mu
},\;\gamma \eta _a^{*}=-f_{\;\;ac}^b\eta _b^{*}\eta ^c, 
\end{equation}
\begin{equation}
\label{21}\delta \eta ^{*\mu }=-2\partial _\nu B^{*\nu \mu },\;\gamma \eta
^{*\mu }=0, 
\end{equation}
\begin{equation}
\label{22}\delta \eta ^{*}=\partial _\mu \eta ^{*\mu },\;\gamma \eta ^{*}=0, 
\end{equation}
where $\left( D_\mu \right) _a^{\;\;b}=\delta _a^{\;\;b}\partial _\mu
-f_{\;\;ac}^bA_\mu ^c$. Formulas (\ref{13}--\ref{22}) will be useful in the
next section at deforming the solution (\ref{5}) by means of cohomological
arguments.

\section{Deformed solution of the master equation}

Here, we will use the technique of deformation of the master equation with
respect to the ``free'' theory, and will deduce the consistent interactions
that can be introduced between the Yang-Mills vector potential and the
2-form.

A consistent deformation of action (\ref{1}) and of its gauge invariances
defines a deformation of the corresponding solution to the master equation
that preserves both the master equation and field/antifield spectra. So, if 
\begin{equation}
\label{23}S_0^L\left[ A_\mu ^a,B_{\mu \nu }\right] +g\int d^Dx\alpha
_0+g^2\int d^Dx\beta _0+O\left( g^3\right) , 
\end{equation}
stands for a consistent deformation of action (\ref{1}), with deformed gauge
transformations 
\begin{equation}
\label{24}\bar \delta _\epsilon A_\mu ^a=\left( D_\mu \right)
_{\;\;b}^a\epsilon ^b+g\sigma _\mu ^a+O\left( g^2\right) , 
\end{equation}
\begin{equation}
\label{25}\bar \delta _\epsilon B_{\mu \nu }=\partial _{\left[ \mu \right.
}\epsilon _{\left. \nu \right] }+g\xi _{\mu \nu }+O\left( g^2\right) , 
\end{equation}
then the deformed solution to the master equation 
\begin{equation}
\label{26}\bar S=S+g\int d^Dx\alpha +g^2\int d^Dx\beta +O\left( g^3\right)
=S+gS_1+g^2S_2+O\left( g^3\right) , 
\end{equation}
satisfies 
\begin{equation}
\label{27}\left( \bar S,\bar S\right) =0, 
\end{equation}
where 
\begin{equation}
\label{28}\alpha =\alpha _0+A_a^{*\mu }\bar \sigma _\mu ^a+B^{*\mu \nu }\bar
\xi _{\mu \nu }+{\rm ``more"}. 
\end{equation}

The master equation (\ref{27}) splits according to the deformation parameter 
$g$ as 
\begin{equation}
\label{29}s\alpha =\partial _\mu k^\mu , 
\end{equation}
\begin{equation}
\label{30}s\beta +\frac 12\omega =\partial _\mu \theta ^\mu , 
\end{equation}
$$
\vdots 
$$
for some local $k^\mu $ and $\theta ^\mu $, with 
\begin{equation}
\label{31}\left( S_1,S_1\right) =\int d^Dx\omega . 
\end{equation}
Obviously, equation (\ref{27}) is automatically satisfied at order zero in
the coupling constant, and has thus been omitted. Equation (\ref{29}) shows
that the non-trivial first-order consistent interactions belong to $%
H^0\left( s|d\right) $, where $d$ is the exterior space-time derivative. In
the case where $\alpha $ is a coboundary modulo $d$ ($\alpha =s\lambda
+\partial _\mu \pi ^\mu $), then the deformation is trivial (it can be
eliminated by a redefinition of the fields).

In order to solve equation (\ref{29}) we expand $\alpha $ accordingly the
antighost number 
\begin{equation}
\label{32}\alpha =\alpha _0+\alpha _1+\cdots +\alpha _k,\;antigh\left(
\alpha _i\right) =i, 
\end{equation}
where the last term in (\ref{32}) can be assumed to be annihilated by $%
\gamma $. Thus, since $antigh\left( \alpha _k\right) =k$ and $gh\left(
\alpha _k\right) =0$, the pure ghost number of $\alpha _k$ is equal to $k$.
The fact that 
\begin{equation}
\label{33}\rho =\frac 13f_{abc}\eta ^a\eta ^b\eta ^c, 
\end{equation}
is $\gamma $-invariant enforces that $k=3m>0$. Then, 
\begin{equation}
\label{34}\alpha _k=\alpha _{3m}=\mu _{3m}\left( \rho \right) ^m, 
\end{equation}
where $\mu _{3m}$ belongs to $H_{3m}\left( \delta |d\right) $. As in the
case of the theory under study $H_j\left( \delta |d\right) =0$ for $j>3$ 
\cite{20}, it follows that the last term in (\ref{32}) has the form 
\begin{equation}
\label{35}\alpha _3=\frac 13\mu _3f_{abc}\eta ^a\eta ^b\eta ^c, 
\end{equation}
with $\mu _3$ from $H_3\left( \delta |d\right) $, hence solution to the
equation 
\begin{equation}
\label{36}\delta \mu _3+\partial _\mu j^\mu =0, 
\end{equation}
for some $j^\mu $. From the former relation in (\ref{22}) we find that the
general representative of $H_3\left( \delta |d\right) $ is 
\begin{equation}
\label{37}\mu _3=\eta ^{*}, 
\end{equation}
therefore 
\begin{equation}
\label{38}\alpha _3=\frac 13f_{abc}\eta ^{*}\eta ^a\eta ^b\eta ^c. 
\end{equation}
At antighost number two, equation (\ref{29}) takes the form 
\begin{equation}
\label{39}\delta \alpha _3+\gamma \alpha _2=\partial _\mu v^\mu , 
\end{equation}
for some $v^\mu $. As 
\begin{equation}
\label{40}\delta \alpha _3=-\frac 13f_{abc}\left( \partial _\mu \eta ^{*\mu
}\right) \eta ^a\eta ^b\eta ^c, 
\end{equation}
from (\ref{39}) we get that 
\begin{equation}
\label{41}\alpha _2=-f_{abc}\eta ^{*\mu }\eta ^a\eta ^bA_\mu ^c, 
\end{equation}
hence 
\begin{equation}
\label{42}\delta \alpha _3+\gamma \alpha _2=\partial _\mu \left(
-f_{abc}\eta ^{*\mu }\eta ^a\eta ^b\eta ^c\right) . 
\end{equation}
By projecting (\ref{29}) on antighost number one, we infer 
\begin{equation}
\label{43}\delta \alpha _2+\gamma \alpha _1=\partial _\mu u^\mu . 
\end{equation}
Starting from 
\begin{equation}
\label{44}\delta \alpha _2=2f_{abc}\left( \partial _\nu B^{*\nu \mu }\right)
\eta ^a\eta ^bA_\mu ^c, 
\end{equation}
we arrive at 
\begin{equation}
\label{45}\alpha _1=2B^{*\mu \nu }\eta _a\partial _{\left[ \mu \right.
}A_{\left. \nu \right] }^a, 
\end{equation}
which leads to 
\begin{equation}
\label{46}\delta \alpha _2+\gamma \alpha _1=\partial _\mu \left(
2f_{abc}B^{*\mu \nu }\eta ^a\eta ^bA_\nu ^c\right) . 
\end{equation}
With $\alpha _1$ at hand, we proceed to determine $\alpha _0$ as solution to
the equation 
\begin{equation}
\label{47}\delta \alpha _1+\gamma \alpha _0=\partial _\mu w^\mu . 
\end{equation}
On account of (\ref{45}), we have 
\begin{equation}
\label{48}\delta \alpha _1=\left( \partial _\rho F^{\rho \mu \nu }\right)
\eta _a\partial _{\left[ \mu \right. }A_{\left. \nu \right] }^a. 
\end{equation}
Then, 
\begin{equation}
\label{49}\alpha _0=\frac 13F^{\mu \nu \rho }\left( f_{abc}A_\mu ^aA_\nu
^bA_\rho ^c+A_\mu ^aF_{a\nu \rho }+A_\rho ^aF_{a\mu \nu }+A_\nu ^aF_{a\rho
\mu }\right) , 
\end{equation}
such that 
\begin{equation}
\label{50}\delta \alpha _1+\gamma \alpha _0=\partial _\mu \left( F^{\mu \nu
\rho }\eta _a\partial _{\left[ \nu \right. }A_{\left. \rho \right]
}^a\right) . 
\end{equation}
In this way, we have generated the first-order deformation under the form%
\begin{eqnarray}\label{51}
& &S_1=\int d^Dx\left( \frac 13F^{\mu \nu \rho }\left( f_{abc}A_\mu ^aA_\nu
^bA_\rho ^c+A_\mu ^aF_{a\nu \rho }+A_\rho ^aF_{a\mu \nu }+A_\nu ^aF_{a\rho
\mu }\right) +\right. \nonumber \\ 
& &\left. 2B^{*\mu \nu }\eta _a\partial _{\left[ \mu \right.
}A_{\left. \nu \right] }^a-f_{abc}\eta ^{*\mu }\eta ^a\eta ^bA_\mu ^c+\frac
13f_{abc}\eta ^{*}\eta ^a\eta ^b\eta ^c\right) . 
\end{eqnarray}

The existence of $\alpha $ is equivalent to the consistency of the
interaction up to order $g$. The interaction is then consistent also to
order $g^2$ if and only if $\omega $ is $s$-exact modulo $d$ (see (\ref{30}%
)). By direct computation, we deduce%
\begin{eqnarray}\label{52}
& &\left( S_1,S_1\right) =
\frac 43\int d^Dx\left( f_{abc}A_\mu ^aA_\nu ^bA_\rho
^c+A_\mu ^aF_{a\nu \rho }+A_\rho ^aF_{a\mu \nu }+\right. \nonumber \\ 
& &\left. \left. A_\nu ^aF_{a\rho \mu }\right) 
\left( \left( \partial ^\mu \eta
_d\right) \partial ^{\left[ \nu \right. }A^{\left. \rho \right] d}+\left(
\partial ^\rho \eta _d\right) \partial ^{\left[ \mu \right. }A^{\left. \nu
\right] d}+\left( \partial ^\nu \eta _d\right) \partial ^{\left[ \rho
\right. }A^{\left. \mu \right] d}\right) \right) \nonumber \\ 
& &=\int d^Dx\omega .
\end{eqnarray}
From the obvious relations 
\begin{eqnarray}\label{53}
& &s\left( f_{abc}A_\mu ^aA_\nu ^bA_\rho ^c+A_\mu ^aF_{a\nu \rho }+A_\rho
^aF_{a\mu \nu }+A_\nu ^aF_{a\rho \mu }\right) =\nonumber \\ 
& &\left( \partial _\mu \eta _d\right) \partial _{\left[ \nu \right.
}A_{\left. \rho \right] }^d+\left( \partial _\rho \eta _d\right) \partial
_{\left[ \mu \right. }A_{\left. \nu \right] }^d+\left( \partial _\nu \eta
_d\right) \partial _{\left[ \rho \right. }A_{\left. \mu \right] }^d,
\end{eqnarray}
we derive that 
\begin{equation}
\label{54}\omega =s\left( \frac 23\left( f_{abc}A_\mu ^aA_\nu ^bA_\rho
^c+A_\mu ^aF_{a\nu \rho }+A_\rho ^aF_{a\mu \nu }+A_\nu ^aF_{a\rho \mu
}\right) ^2\right) ,
\end{equation}
which further yields 
\begin{equation}
\label{55}S_2=-\frac 13\int d^Dx\left( f_{abc}A_\mu ^aA_\nu ^bA_\rho
^c+A_\mu ^aF_{a\nu \rho }+A_\rho ^aF_{a\mu \nu }+A_\nu ^aF_{a\rho \mu
}\right) ^2.
\end{equation}
The higher-order equations are then satisfied with 
\begin{equation}
\label{56}S_3=S_4=\cdots =0.
\end{equation}
In this way, we inferred that%
\begin{eqnarray}\label{57}
& &\bar S=\int d^Dx\left( -\frac 14F_{\mu \nu }^aF_a^{\mu \nu }-\frac
1{12}H_{\mu \nu \rho }H^{\mu \nu \rho }+A_a^{*\mu }\left( D_\mu \right)
_{\;\;b}^a\eta ^b-\right. \nonumber \\ 
& &\frac 12f_{\;\;bc}^a\eta _a^{*}\eta ^b\eta ^c+
B^{*\mu \nu }\left( \partial
_{\left[ \mu \right. }\eta _{\left. \nu \right] }+2g\eta _a\partial _{\left[
\mu \right. }A_{\left. \nu \right] }^a\right) +\eta ^{*\mu }\partial _\mu
\eta -\nonumber \\ 
& &\left. gf_{abc}\eta ^{*\mu }\eta ^a\eta ^bA_\mu ^c+\frac
g3f_{abc}\eta ^{*}\eta ^a\eta ^b\eta ^c\right) ,
\end{eqnarray}
is solution to the master equation (\ref{27}) of our deformation problem.
Clearly, solution (\ref{57}) is covariant and local.  The
antifield-independent piece in (\ref{57}) 
\begin{equation}
\label{58}\bar S_0=\int d^Dx\left( -\frac 14F_{\mu \nu }^aF_a^{\mu \nu
}-\frac 1{12}H_{\mu \nu \rho }H^{\mu \nu \rho }\right) ,
\end{equation}
with 
\begin{equation}
\label{59}H_{\mu \nu \rho }=F_{\mu \nu \rho }-2g\left( f_{abc}A_\mu ^aA_\nu
^bA_\rho ^c+A_{\left[ \mu \right. }^aF_{a\left. \nu \rho \right] }\right) ,
\end{equation}
describes the combined Yang-Mills-2-form system coupled through the
Yang-Mills Chern-Simons term, known as the Chapline-Manton model. From the
terms linear in the antifields of the original fields we find that the gauge
transformations of the Yang-Mills vector potential are not deformed, while
those of the 2-form are so, namely, 
\begin{equation}
\label{60}\bar \delta _\epsilon B_{\mu \nu }=\partial _{\left[ \mu \right.
}\epsilon _{\left. \nu \right] }+2g\epsilon _a\partial _{\left[ \mu \right.
}A_{\left. \nu \right] }^a.
\end{equation}
The analysis of the terms linear in the antifields of the ghosts emphasizes
that the commutators among the gauge generators of the Yang-Mills fields
remain unaffected, but there appear non-vanishing commutators between the
gauge generators of the 2-form and Yang-Mills fields (see the supplementary
term $-gf_{abc}\eta ^{*\mu }\eta ^a\eta ^bA_\mu ^c$). Thus, the deformation
problem analyzed here maintains the covariance and space-time locality, and
leads to the deformation of both gauge transformations and gauge algebra.
This completes our approach.

\section{Conclusion}

To conclude with, in this paper we have investigated the consistent
interaction that can be introduced between the Yang-Mills vector potential
and a 2-form. Starting with the BRST differential for the ``free'' theory, $%
s=\delta +\gamma $, we initially compute $H^0\left( s|d\right) $ by
expanding the co-cycles accordingly the antighost number, and generate in
this way the consistent first-order deformation. Next, we prove that the
deformation is also second-order consistent and, moreover, matches the
higher-order deformation equations. As a result, we are led precisely to the
Chern-Simons couplings of a 2-form to Yang-Mills gauge fields, that imply
the deformation of both gauge transformations and their algebra. Finally, we
remark that in the case of the coupled model the ghost for ghost $\eta $ and
the primitive form $f_{abc}\eta ^a\eta ^b\eta ^c$ do not enter the
cohomology since $f_{abc}\eta ^a\eta ^b\eta ^c=\bar s\left( \frac 3g\eta
\right) $, which indicates that the primitive form is $\bar s$-exact and $%
\eta $ is no longer closed. In turn, this last property underlies the
Green-Schwarz anomaly cancellation mechanism.

\end{document}